\documentclass[aps,prl,showpacs,showkeys,twocolumn,groupedaddress,superscriptaddress]{revtex4}
\usepackage{graphicx}%
\usepackage{amsmath}
\usepackage{dcolumn}

\usepackage{graphicx}%
\usepackage{dcolumn}

\begin{document}

\title{Observation of an Inverse Energy Cascade\\ in Developed
Acoustic Turbulence in Superfluid Helium}

\author{A.~N.~Ganshin}%
\affiliation{Department of Physics, Lancaster University,
Lancaster, LA1 4YB, UK}

\author{V.~B.~Efimov}%
\affiliation{Department of Physics, Lancaster University,
Lancaster, LA1 4YB, UK}
\affiliation{Institute of Solid State Physics RAS, Chernogolovka,
Moscow region, 142432, Russia}

\author{G.~V.~Kolmakov\footnote{Currently at the Department of Chemical and Petroleum
Engineering, Pittsburgh University,  Pittsburgh, PA 15261.}}%
\affiliation{Institute of Solid State Physics RAS, Chernogolovka,
Moscow region, 142432, Russia}
\affiliation{Department of Physics, Lancaster University,
Lancaster, LA1 4YB, UK}

\author{L.~P.~Mezhov-Deglin}%
\affiliation{Institute of Solid State Physics RAS, Chernogolovka,
Moscow region, 142432, Russia}

\author{P.~V.~E.~McClintock}%
\affiliation{Department of Physics, Lancaster University,
Lancaster, LA1 4YB, UK}

\date{\today}

\begin{abstract}

We report observation of an inverse energy cascade in second
sound acoustic turbulence in He\,II. Its onset occurs above a
critical driving energy and it is accompanied by giant waves
that constitute an acoustic analogue of the rogue waves that
occasionally appear on the surface of the ocean. The theory of
the phenomenon is developed and shown to be in good agreement
with the experiments.

\end{abstract}



\maketitle

A highly excited state of a system with numerous degrees of freedom,
characterized by a directional energy flux through frequency scales, is
referred to as {\it turbulent} \cite{K41,book}. Like the familiar
manifestations of vortex turbulence in fluids,
turbulence can also occur in systems of waves, e.g.\  turbulence of sound waves
in oceanic waveguides \cite{at1}, magnetic turbulence in interstellar gases
\cite{at2}, shock waves in the solar wind and their coupling with Earth's
magnetosphere \cite{at3}, and phonon turbulence in solids \cite{Tsoi}.
Following the ideas of Kolmogorov, the universally accepted picture says that
nonlinear wave interactions give rise to a cascade of wave energy towards
shorter and shorter wavelengths until, eventually, it becomes possible for
viscosity to dissipate the energy as heat. Experiments and calculations show
that, most of the time, the Kolmogorov picture is correct
\cite{book,PRL2006,falk1}.

We demonstrate below that this picture is incomplete. Our experiments with
second sound (temperature-entropy) waves in He\,II show that, contrary to the
conventional wisdom, acoustic wave energy can sometimes flow in the opposite
direction too. We note that inverse energy cascades are known in 2-dimensional
incompressible liquids and Bose gases \cite{Kraichnan}, and have been
considered for quantized vortices \cite{Vinen}.

We find that energy backflow in our acoustic system is attributable to a decay
instability (cf.\ the kinetic instability in turbulent systems \cite{Lvovkin}),
controlled mainly by nonlinear decay of the wave into two waves of lower
frequency governed by the energy (frequency) conservation law \cite{book}
$\omega_1=\omega_2+\omega_3$.
Here $\omega_i=u_{20} k_i$ is the frequency of a linear wave of wave vector
$k_i$ and $u_{20}$ is the second sound velocity at negligibly small amplitude.
The instability manifests itself in the generation of subharmonics.
A quite similar parametric process, due to 4-wave scattering (modulation
instability), is thought to be responsible for the generation of large
wind-driven ocean waves \cite{Diachenko}. Decay instabilities (especially
threshold and near-threshold behaviour) have been studied for e.g.\ spin waves
\cite{Suhl}, magnetohydrodynamic waves in plasma \cite{Spangler}, and
interacting first and second sound waves in superfluid helium near the
superfluid transition \cite{Rinberg}.

We now discuss what happens to a system of acoustic waves far beyond the decay
threshold. Modelling the resultant nonlinear wave transformations in the
laboratory is a potentially fruitful approach that has already yielded
important results for e.g.\ the turbulent decay of capillary waves on the
surface of liquid H$_2$ \cite{prl}. Here, we exploit the special properties of
second sound \cite{Landau}, which enable fundamental wave processes to be
studied under laboratory conditions. Its velocity $u_2$ depends strongly on its
amplitude $\delta T$ and can be approximated as
\begin{equation}
u_2 = u_{20} (1 + \alpha \delta T),
\end{equation}
where the nonlinearity coefficient $\alpha$ \cite{Fairbank},
which determines the strength of the wave interactions, can be
made large if the temperature is set right.

\begin{figure}[t]
\vspace*{-0.5cm}
\centerline{\includegraphics[width=87.mm]
{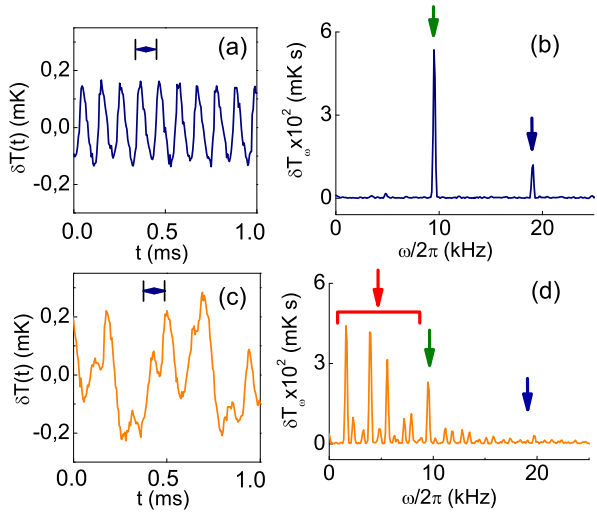}
} \vspace{-0.5cm} \caption{\footnotesize (Color online) Evolution
of the observed wave shape in the resonator (left column) and of
the power spectrum of second sound standing waves (right) with
increasing drive frequency $\omega_d$ near the 96th resonance:
$\omega_d/2 \pi=9530.8$ Hz (a),(b);
and 9535.2 Hz (c),(d). The AC heat flux density was $W=42$
mW/cm$^2$. The temperature $T=2.08$ K corresponded to negative nonlinearity.
The fundamental and first harmonic in (b),(d) are indicated by vertical
(green and blue) arrows; the low-frequency domain where the subharmonics appear
are indicated by the horizontal range in (d) with vertical (red) arrows.
The horizontal arrows in (a),(c) indicate the fundamental period of a wave
at the driving frequency.} \label{fig2}
\end{figure}

Our experiments make use of the high $Q$ cryoacoustical resonator
described previously \cite{PRL2006}, excited close to one of its
resonant frequencies. It enables very large second sound
{standing} wave amplitudes to be attained for modest levels of
excitation.
Its $Q$-factor, determined from the widths of longitudinal
resonances at small heat flux densities, was $Q \sim 1000$ for
resonant numbers $p \leq 10$ and $Q \sim 3000$ for $30<p<100$.
Fig.\ \ref{fig2} presents typical results obtained when driving at
a relatively high resonant frequency $\omega_d$ (the 96th
longitudinal resonance of the cell). Those in Figs.\
\ref{fig2}(a),(b) reproduce our earlier observation of the direct
Kolmogorov-like cascade of second sound waves in He~II
\cite{PRL2006}, when driving on resonance. Figs.\
\ref{fig2}(c)--(d) show that tiny changes in driving frequency can
produce marked changes in the shape and spectrum of the standing
wave. The formation of the spectral peaks near $\frac{1}{2}$,
$\frac{1}{3}$ and $\frac{2}{3}$ $\times$ $\omega_{d}$ satisfies
the frequency conservation law with $\omega_1=\omega_d$,
supporting our inference that the instability is controlled mainly
by 3-wave interactions. When the instability develops, huge
distortions of the initially periodic signal occur. Although it
remains nearly periodic, its characteristic period exceeds that of
the driving force and its amplitude can become more than twice
that at the driving frequency.

\begin{figure}[!t]
\includegraphics [width=75.mm]{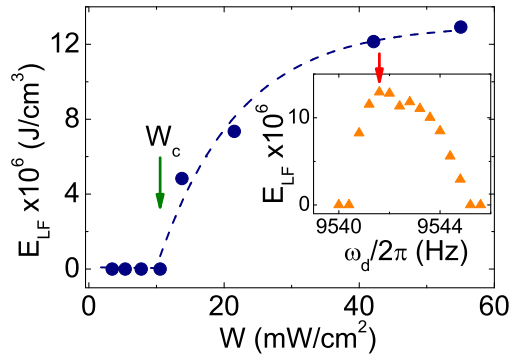}
\vspace{-0.5cm} \caption{\footnotesize (Color online) The energy
$E_{LF}$ contained in the low-frequency part of the spectrum as a
function of the AC heat flux density $W$, while driving near to
the 96th resonance for $T \simeq 2.08$~K. The threshold value of
$W$, marked by the (green) arrow, was $W_c=10.4$ mW/cm$^2$. The
points are from experiment; dashed lines are guides to the eye.
Inset: the dependence of $E_{LF}$ on $\omega_d$, measured for
$W=55.6$~mW/cm$^2$; the (red) arrow labels the maximum value of
$E_{LF}$, which is taken to the main figure.} \label{fig3}
\end{figure}

To characterise the instability quantitatively, we use the
energy contained in the low-frequency part of the spectrum
$\omega<\omega_d$,
\begin{equation}
E_{LF} = {1 \over 2} \left({\partial C \over \partial T}\right)
\sum_{ \omega<\omega_d} | \delta T_{\omega}|^2,
\end{equation}
as an indicator. Fig.\ \ref{fig3} shows the dependence of
$E_{LF}$ on the AC heat flux density $W$, when driving on the
96th resonance close to 2.08 K. For small $W$ we did not
observe any subharmonic generation at all \cite{PRL2006}. Then,
above a critical flux $W_c$, $E_{LF}$ rose rapidly, suggesting
that the phenomenon is of a threshold character. At
$T=T_{\alpha}=1.88$~K for which $\alpha$ vanishes
\cite{Fairbank}, no subharmonics were observed, regardless of
the magnitude of $W$, thus confirming the crucial importance of
nonlinearity. For $W$ above 10.4 mW/cm$^2$, we observed a
distortion of the signal similar to that shown in Fig.\
\ref{fig2}(c) and the formation of a few subharmonics. Further
increase of $W$ above 20 mW/cm$^2$ led to the generation of
multiple subharmonics. These phenomena appear in the regime
where the energy cascade towards the high frequency domain
(i.e.\ direct cascade, with a Kolmogorov-like spectrum
\cite{PRL2006,falk1}), is already well-developed; see also
Fig.\ \ref{figst}.

\begin{figure}[t]
\vspace*{-0.5cm}
\begin{center}
\includegraphics[width=75mm]{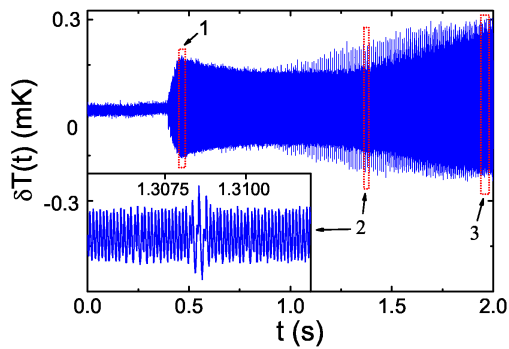}
\end{center}
\vspace{-1.0cm} \caption{(Color online) Transient evolution of the
2nd sound wave amplitude $\delta T$ after a step-like shift of the
driving frequency to the 96th resonance at time $t=0.397$\,s.
Signals in frames 1 and 3 are similar to those obtained in
steady-state measurements, Fig. 1 (a) and (e) respectively.
Formation of isolated ``rogue'' waves are clearly evident. Inset:
example of a rogue wave, enlarged from frame 2.} \label{newfig3}
\end{figure}

\begin{figure}[t]
\vspace*{-0.5cm}
\begin{center}
\includegraphics[width=70mm]{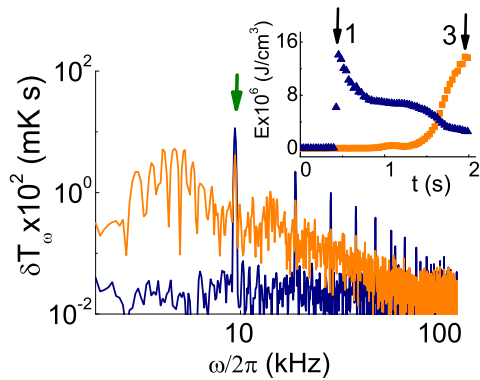}
\end{center}
\vspace{-1.0cm} \caption{(Color online) Instantaneous spectra in
frames 1 and 3 of Fig. \ref{newfig3}. The lower (blue) spectrum,
for frame 1, shows the direct cascade only; the upper (orange)
spectrum, for frame 3, shows both the direct and inverse cascades.
The (green) arrow indicates the the fundamental peak at the
driving frequency. Inset: evolution of the wave energy in the
low-frequency and high frequency domains is shown by the (orange)
squares and (blue) triangles respectively; (black) arrows mark the
positions of frames 1 and 3.} \label{figst}
\end{figure}

All the above results correspond to {\it steady-state} regimes, of the wave
system. In Fig.\ \ref{newfig3}  we illustrate the {\it transient} processes
observed after a step-like shift of the driving frequency from a frequency
initially set far from any resonance to the 96th resonance frequency for
$W=42.1$\,mW/cm$^2$, $T=2.08$ K. We find that harmonics of the drive in the
high-frequency spectral domain are formed very quickly, but that formation of
the subharmonics takes much longer: it took $\sim 0.5$ s here, and can reach
several tens of seconds under some conditions \cite{PREWT}. It is evident from
the inset in Fig. \ref{newfig3} that, as the instability develops, isolated
``rogue waves'' appear in the signal. As time evolves, the rogue waves appear
more frequently and, at the later stages, they merge resulting in the strong
low-frequency modulation of the signal observed in the steady-state
measurements (Fig.\ref{fig2}).

\begin{figure}[t]
\vspace*{-1.0cm}
\centerline{\includegraphics[width=70.mm]{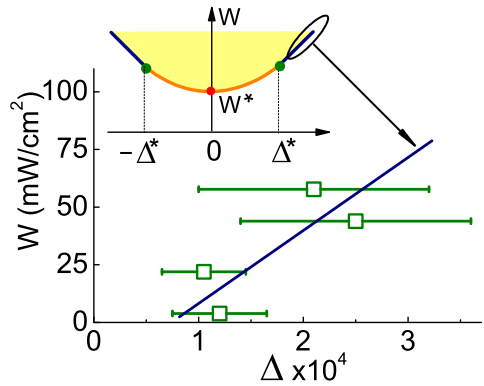}}
\caption{(Color online) Dependence of the AC heat flux density $W$
at which the instability develops on the dimensionless frequency
detuning $\Delta = (\omega_d - \omega_n)/\omega_n$ of the driving
force frequency $\omega_d$ from a cavity resonance $\omega_n$.
Numerical calculations (line) are compared with measurements
(points) for driving at the 96th resonance. Horizontal bars mark
the widths of the hysteretic region where second sound
exists in a metastable state. Inset: bifurcation diagram showing
regions of stability (unshaded) and and regions of
instability (yellow shaded) against the generation of subharmonics. The
soft instability occurs over the (orange) line between the (green)
critical points at $\pm \Delta^{\ast}$; outside them lies the hard
instability; $W^*$ is the threshold value of the
instability.} \label{fig1}
\end{figure}

When the subharmonics appear, a marked reduction occurs in the energy contained
in the high-frequency spectral domain: see Fig. \ref{figst}. A substantial
proportion of the energy then flows from the driving frequency $\omega_d$
towards the low frequency domain $\omega < \omega_d$ leading to an accumulation
of wave energy there and a corresponding increase in wave amplitude. The
reduction of wave amplitude seen in the high frequency spectral domain is
indicative of the onset of energy backflow towards lower frequencies, i.e.\ a
sharing of the flux between the direct and inverse energy cascades. The
decrease in energy at high frequencies energy in $0.397 s < t < 1.3$\,s is
attributable to relaxation processes in the direct cascade. Redistribution of
wave energy due to sharing of the energy flux between the direct and inverse
cascades starts at $t=1.3$\,s. Note that the transient evolution shown in Figs
3 and 4 is incomplete: the forward and inverse energy fluxes are still changing
at $t=2s$, leading us to anticipate further relaxation oscillations at longer
times (inaccessible with our present equipment): the transient dynamics is
clearly complex, and a full characterisation will require further work.
Absorption of wave energy at low frequencies is probably attributable to
viscous drag of the normal fluid component on the resonator walls, given that
bulk second sound damping is negligibly small in this frequency range: this
would be consistent with the observed strong decrease of the resonator
$Q$-factor below 3\,kHz. We observe {\it hysteresis} between increasing and
decreasing frequency scans (bars on data points in Fig.\ \ref{fig1}). Its
width, i.e.\ the region where the low frequency sound waves are metastable, was
less than the viscous width of the resonance.

To seek a more detailed understanding of wave energy transformation in acoustic
turbulence, we used a technique \cite{PREWT} for direct numerical integration
of the 2-fluid thermohydrodynamical equations \cite{Landau}, expanded up to
quadratic terms in the wave amplitude. It representats the second sound waves
in terms of Hamiltonian variables \cite{PRL2006,Pokrovsky}, as in earlier
studies \cite{falk1} of acoustic turbulence. Wave damping was, however, taken
explicitly into account at all frequencies, a feature that is of key importance
for a correct description of subharmonic generation. The main results are
summarised in Fig.\ \ref{fig1}.

It is evident (inset of Fig.\ \ref{fig1}) that, for sufficiently high driving
amplitude $W$, the wave develops an instability with respect to generation of
low-frequency subharmonics of the driving force at $\omega_d$. For zero
detuning from a cavity resonance, the onset of the instability occurs at a
threshold $W^{\ast} \propto 1/\alpha$. If the dimensionless frequency detuning
$|\Delta |$ is less than a critical value $\Delta^* \sim 1/Q$, the instability
has a soft character, in that the amplitudes of the low frequency waves tend to
zero at the threshold bifurcation line. Outside this range, the low frequency
waves are characterised by the hard onset observed experimentally. Measurements
(squares) are compared with theory (full line) in the main part of
Fig.~\ref{fig1}. The hard onset is accompanied by a finite jump in subharmonic
amplitude. These two regimes of behaviour are separated by critical points on
the bifurcation line. The generation of subharmonics in the nonlinear
oscillatory system found numerically appears to be similar to the bifurcation
of an anharmonic oscillator \cite{Mandelstam}. We estimated numerically that
the critical detuning parameter $\Delta^*$ at $T=2.08$ K is close to $10^{-4}$,
and that the critical AC heat flux density $W^{\ast}$ is equal to a few
mW/cm$^2$, in good agreement with our observations. The experimental results
shown in Fig.\ \ref{fig2} can be understood as corresponding to the working
point moving horizontally on the bifurcation diagram (Fig.~\ref{fig1}, inset)
into the (yellow) shaded region from a position just outside it, through the
hard (blue) instability line. Although nonlocal (in $\omega$-space)
interactions between waves at $\omega_d$ and low frequency waves probably
contributes to the subharmonics, due to the finite width of the low-frequency
spectral domain, our numerical estimates show that the interactions of
high-frequency ($\omega \gg \omega_d$) waves  with subharmonics are relatively
weak, consistent with our inference of two cascades.

Our experiments have revealed an inverse wave energy cascade in a turbulent
acoustic system. It is responsible for a substantial increase in wave amplitude
corresponding to the formation of a set of huge low-frequency subharmonics.
This instability develops through formation of isolated low-frequency waves of
amplitude higher than that typical of waves around them. The latter can be
considered as the acoustic analogue of the giant ``rogue'' waves that
occasionally appear on the ocean and endanger shipping. Their origin lies in
the development of a decay instability of the periodic wave, i.e.\ a similar
mechanism to that proposed \cite{Onorato,Diachenko} (modulation instability) to
account for the creation of oceanic rogue waves \cite{Dean}. Note that this
mechanism differs from an alternative explanation proposed recently
\cite{natureopt} that involves scattering of nonlinear waves on a continuous
noisy background.

We acknowledge valuable discussions with A. A. Levchenko, V. E.
Zakharov, E. A. Kuznetsov, V. S. L'vov, V. V. Lebedev and T.
Mizusaki. The work was supported by Engineering and Physical
Sciences Research Council (UK), the Russian Foundation for
Basic Research, project No. 07-02-00728 and by the Presidium
of the Russian Academy of Sciences under the programs ``Quantum
Macrophysics'' and ``Mathematical Methods in Nonlinear
Dynamics''.

\end{document}